\documentclass[sigconf, authorversion=true]{acmart}
\copyrightyear{2024}
\acmYear{2024}
\setcopyright{rightsretained}
\acmConference[ICCAD '24]{IEEE/ACM International Conference on Computer-Aided Design}{October 27--31, 2024}{New York, NY, USA}
\acmBooktitle{IEEE/ACM International Conference on Computer-Aided Design (ICCAD '24), October 27--31, 2024, New York, NY, USA}
\acmDOI{10.1145/3676536.3676841}
\acmISBN{979-8-4007-1077-3/24/10}

\acmConference[To appear in ICCAD '24]{International Conference on Computer-Aided Design}{October 27--31,2024}{New Jersey, NJ}

\usepackage{algorithm}

\usepackage{subcaption}

\usepackage{etoolbox}
\usepackage[normalem]{ulem}

\newtoggle{STRIKEMODE}
\togglefalse{STRIKEMODE}
\newcommand{\CITE}[1]{\iftoggle{STRIKEMODE}{\textbackslash CITE\{#1\}}{\cite{#1}}}

\newcommand{\refsec}[1]{Sec.~\ref{sec:#1}}
\newcommand{\reffig}[1]{Fig.~\ref{fig:#1}}

\newcommand{\RISCV}{\mbox{RISC-V}}

\newcommand{\PALERT}{\mbox{P-alert}}

\newcommand{\LALERT}{\mbox{L-alert}}

\PassOptionsToPackage{hyphens}{url}
\usepackage{url}

\usepackage{framed}

\usepackage{amsthm}

\theoremstyle{definition}

\usepackage{listings}
\usepackage{tcolorbox}
\tcbuselibrary{listings,skins}

\definecolor{indigo}{RGB}{51,34,136}
\definecolor{cyan}{RGB}{136,204,238}
\definecolor{teal}{RGB}{68,170,153}
\definecolor{green}{RGB}{17,119,51}
\definecolor{olive}{RGB}{153,153,51}
\definecolor{sand}{RGB}{221,204,119}
\definecolor{rose}{RGB}{204,102,119}
\definecolor{wine}{RGB}{136,34,85}
\definecolor{purple}{RGB}{170,68,153}
\definecolor{palegrey}{RGB}{221,221,221}
\definecolor{codegreen}{rgb}{0,0.6,0}
\definecolor{codegray}{rgb}{0.5,0.5,0.5}
\definecolor{codepurple}{rgb}{0.58,0,0.82}
\definecolor{backcolour}{rgb}{0.95,0.95,0.92}
\definecolor{onespin}{RGB}{95,0,142}

\lstdefinestyle{assembler}{
    frame=single,
    numbers=left,
    numbersep=5pt,
    captionpos=b,
    mathescape=true,
    tabsize=2,
    breaklines=true,
    breakatwhitespace=false,
    keepspaces=true,
    showspaces=false,
    showstringspaces=false,
    showtabs=false,
    backgroundcolor=\color{backcolour},
    commentstyle=\color{green},
    keywordstyle=\color{blue},
    numberstyle=\tiny\color{indigo},
    stringstyle=\color{purple},
    basicstyle=\ttfamily\footnotesize,
    moredelim=**[is][\btHL]{@}{@},
    keywordstyle=[2]{\color{onespin}},
    keywordstyle=[3]{\color{blue}},
    keywordstyle=[4]{\color{purple}},
    otherkeywords={+,*,;, :, |, &, 1'b0, 1'b1, jmp, jg, cmp},
    morekeywords=[2]{+,*, ;, :, =, |, &},
    morekeywords=[3]{{1`b0}, {1`b1}, mov,[,]},
    morekeywords=[4]{rax, rbx, rcx, rdx, r8, r9, r10},
    xleftmargin=1.5em,
    framexleftmargin=1em
}

\lstdefinestyle{assembler2}{
    frame=none,
    numbers=left,
    numbersep=5pt,
    captionpos=b,
    mathescape=true,
    tabsize=2,
    breaklines=true,
    breakatwhitespace=false,
    keepspaces=true,
    showspaces=false,
    showstringspaces=false,
    showtabs=false,
    commentstyle=\color{green},
    keywordstyle=\color{blue},
    numberstyle=\tiny\color{indigo},
    stringstyle=\color{purple},
    basicstyle=\ttfamily\footnotesize,
    moredelim=**[is][\btHL]{@}{@},
    keywordstyle=[2]{\color{onespin}},
    keywordstyle=[3]{\color{blue}},
    keywordstyle=[4]{\color{purple}},
    otherkeywords={+,*,;, :, |, &, 1'b0, 1'b1, jmp, jg, cmp},
    morekeywords=[2]{+,*, ;, :, =, |, &},
    morekeywords=[3]{{1`b0}, {1`b1}, mov,[,]},
    morekeywords=[4]{rax, rbx, rcx, rdx, r8, r9, r10},
    xleftmargin=1.5em,
    framexleftmargin=1em
}

\lstdefinestyle{onespin}{
    backgroundcolor=\color{backcolour},   
    commentstyle=\color{codegreen},
    keywordstyle=\color{magenta},
    numberstyle=\tiny\color{codegray},
    stringstyle=\color{codepurple},
    basicstyle=\ttfamily\footnotesize,
    moredelim=**[is][\btHL]{@}{@},
    breakatwhitespace=false,         
    breaklines=true,                 
    captionpos=b,                    
    keepspaces=true,                 
    numbers=left,                    
    numbersep=5pt,                  
    showspaces=false,                
    showstringspaces=false,
    showtabs=false,                  
    tabsize=2,
    keywordstyle = [2]{\color{onespin}},
    keywordstyle = [3]{\color{blue}},    
    otherkeywords = {[, ], +, ;, :, =, |, &, !, 1'b0, 1'b1},
    morekeywords = [2]{[, ], +, ;, :, =, |, &, !}
    morekeywords = [3]{{1`b0}, {1`b1}}
}

\newtcblisting{property}[2][]{  
    arc=0pt, outer arc=0pt,
    listing only, 
    listing style=assembler2,
    title=#2,
    colback=backcolour, 
    boxrule=0.5pt,
    boxsep=-5pt,
    left=4pt,
    #1
    }

\begin{document}

\newcommand{\TitleFirst}{%
  VeriCHERI: \\Exhaustive Formal Security Verification of CHERI
  at the RTL
}

\newcommand{\TitleSecond}{%
}
\title{%
\TitleFirst%
\\%
\TitleSecond%
}

\author{Anna Lena Duque Ant\'{o}n}
\authornote{Both authors contributed equally to this research.}
\affiliation{%
	\institution{RPTU Kaiserslautern-Landau}
	\city{Kaiserslautern}%
	\country{Germany}%
}
\email{anna.duqueanton@rptu.de}

\author{Johannes M\"{u}ller}
\authornotemark[1]
\affiliation{%
	\institution{RPTU Kaiserslautern-Landau}
	\city{Kaiserslautern}%
	\country{Germany}%
}
\email{johannes.mueller@rptu.de}

\author{Philipp Schmitz}
\affiliation{%
	\institution{RPTU Kaiserslautern-Landau}
	\city{Kaiserslautern}%
	\country{Germany}%
}
\email{schmitzp@rptu.de}

\author{Tobias Jauch}
\affiliation{%
	\institution{RPTU Kaiserslautern-Landau}
	\city{Kaiserslautern}%
	\country{Germany}%
}
\email{tobias.jauch@rptu.de}

\author{Alex Wezel}
\affiliation{%
	\institution{RPTU Kaiserslautern-Landau}
	\city{Kaiserslautern}%
	\country{Germany}%
}
\email{wezel@rptu.de}

\author{Lucas Deutschmann}
\affiliation{%
	\institution{RPTU Kaiserslautern-Landau}
	\city{Kaiserslautern}%
	\country{Germany}%
}
\email{lucas.deutschmann@rptu.de}

\author{Mohammad R. Fadiheh}
\affiliation{%
	\institution{Stanford University}
	\city{Stanford}%
	\country{United States}%
}
\email{fadiheh@stanford.edu}

\author{Dominik Stoffel}
\affiliation{%
	\institution{RPTU Kaiserslautern-Landau}
	\city{Kaiserslautern}%
	\country{Germany}%
}
\email{dominik.stoffel@rptu.de}

\author{Wolfgang Kunz}
\affiliation{%
	\institution{RPTU Kaiserslautern-Landau}
	\city{Kaiserslautern}%
	\country{Germany}%
}
\email{wolfgang.kunz@rptu.de}

\renewcommand{\shortauthors}{A. Duque Ant\'{o}n, J. M\"{u}ller et al.}

\begin{abstract}
    Protecting data in memory from attackers continues to be a concern
    in computing systems. CHERI is a promising approach to achieve such
    protection, by providing and enforcing fine-grained memory
    protection directly in the hardware. Creating trust for the entire
    system stack, however, requires a gap-free verification of CHERI's
    hardware-based protection mechanisms. Existing verification methods
    for CHERI target the abstract ISA model rather than the underlying
    hardware implementation. Fully ensuring the CHERI security
    guarantees for a concrete RTL implementation is a challenge in
    previous flows and demands high manual efforts.
  
    This paper presents VeriCHERI, a novel approach to security
    verification. It is conceptionally different from previous works in
    that it does not require any ISA specification. Instead of checking
    compliance with a golden ISA model, we check against
    well-established global security objectives of confidentiality and
    integrity. %
    Fully covering these objectives, VeriCHERI uses as few as four
    unbounded properties to exhaustively prove or disprove any
    vulnerability. We demonstrate the effectiveness and scalability of
    VeriCHERI on a RISC-V based processor implementing a CHERI variant.
  \end{abstract}
  
  \keywords{Memory Isolation, Formal Verification, Hardware Security}
  
  \maketitle

  \section{Introduction}
  \label{sec:introduction}
  
  In a world full of complex computing systems, rigid and trustworthy
  security mechanisms are in high demand. %
  A promising contribution to making systems more secure is proposed by
  \emph{Capability Hardware Enhanced RISC Instructions}
  (CHERI)~\cite{2014-WoodruffWatson.etal}. %
  CHERI is designed to enhance RISC instruction set architectures (ISAs) with fine-grained memory
  protection directly implemented and enforced in the hardware (HW). %
  Its architectural concept enables the memory isolation infrastructure
  to enforce memory
  compartmentalization~\CITE{2022-SartakovVilanova.etal,
    2015-WatsonWoodruff.etal}. %
  
  Establishing trust in the proposed mechanisms is particularly
  important for security-focused solutions, and the path to build this trust is exhaustive verification. %
  
  In previous work, considerable effort has been put into formal
  verification of CHERI at the ISA level. %
  In~\CITE{2020-NienhuisJoannou.etal, 2023-GrisenthwaiteBarnes.etal} an
  HOL model of the CHERI ISA based on
  SAIL~\cite{2019-ArmstrongBauereiss.etal} is subjected to security
  verification. %
  Multiple security objectives are formulated and exhaustively proven on
  the SAIL model. %
  This is an important contribution towards establishing trust in the
  security architecture. %
  However, verification results obtained for the ISA level do not
  translate into security guarantees for a HW implementation of the
  ISA. %
  Additional functional verification is required to prove compliance of
  the implementation with the previously verified ISA. %
  
  In an attempt to bridge this gap,~\cite{2021-GaoMelham} proposes to
  verify the CHERI-related subset of the ISA by translating the
  corresponding parts of the SAIL ISA model to SystemVerilog Assertions (SVA) properties to be
  checked against the register transfer level (RTL) implementation of a processor. %
  This is a significant improvement over previous flows. %
  However, even if all security features, as given by the considered ISA
  subset, are proven to be functionally correct, other parts of the
  design, not included in the proofs, may still violate security. %
  Therefore, this approach, in spite of high manual effort, may still
  fall short of guaranteeing security for the design under
  verification. %
  
  Apparently, a \emph{gap-free} formal verification of the processor at
  the RTL is desirable. %
  While methods for so called ``complete''
  ~\cite{2014-UrdahlStoffel.etal} 
  formal
  verification of processors exist and are practiced in industry, for
  many vendors the involved efforts and cost remain, until today,
  prohibitively high. %

  An additional concern is that security verification referring to a time-abstract ISA model misses non-functional
  security vulnerabilities like timing side channels. %
  Consequently, a HW implementation proven to be fully compliant with the ISA model
  may well contain such vulnerabilities. %
  
  In comparison to formal verification procedures carried out for CHERI
  previously, the methodology proposed in this paper, VeriCHERI, uses a
  conceptually different approach. %
  We start with a general, formalized notion of security w.r.t.\
  confidentiality and integrity, and derive unbounded properties from
  it. %
  By no longer deriving properties from the abstract ISA model but
  directly targeting the global security objectives, we eliminate the
  risk of missing security issues. %
  In addition, VeriCHERI detects vulnerabilities that can only be
  observed on the timing-accurate RTL implementation and which are not
  represented on the high-level ISA model. %
  
  Our contributions are, in particular: %
  
  \begin{itemize}
  \item We propose VeriCHERI, a verification framework that exhaustively
    detects all confidentiality and integrity vulnerabilities due to
    functional bugs. %
    In addition, the framework covers all vulnerabilities to
    Meltdown-style attacks. %
    At its core, VeriCHERI consists of only four SVA
    properties, which are checked against the RTL model of the design under verification (DUV). %
  \item VeriCHERI is a ``\emph{spec-free}''
    approach: %
    The security properties are derived from well-established, general
    security objectives rather than from a
    specification or formalized ISA model. %
    Security guarantees are established directly on the RTL model and do
    not have to be imported and rewritten from a
    higher-level model. %
    This has the key advantage of avoiding the
    high costs involved in complete formal verification
    --- and of still being exhaustive w.r.t.\ our
    security goals. %
  \item From the general security objectives we derive scalable properties by
    developing a symbolic representation of all possible tasks potentially
    violating the security objectives. %
    This step significantly improves scalability since it allows us to formulate
    unbounded properties considering only a single execution trace over a finite
    number of clock cycles, as opposed to the security objectives describing infinite
    behavior by comparing two execution traces. Our methodology leads to proofs that are independent of any specific application
    or capability configuration. %
  \item We present a case study on the CHERIoT Ibex core~\CITE{2023-AmarChisnall.etal} and demonstrate
    the efficacy and scalability of VeriCHERI. %
    We detected security bugs including a vulnerability to a potential
      Meltdown-style timing attack. %
      This emphasizes the need for exhaustive security verification
      techniques.
  \end{itemize}
  
  The remainder of the paper provides background on CHERI in Sec.~\ref{sec:cheri} and
  compares the proposed work to related work in Sec.~\ref{sec:related-work}. %
  We then derive a set of properties from our abstract security objective in Sec.~\ref{sec:formal-model}
  and use the properties in the novel VeriCHERI verification flow presented in Sec.~\ref{sec:method}. %
  In Sec.~\ref{sec:case-study}, we present a case study and conclude the paper with Sec.~\ref{sec:conclusion}. %

  \section{Background}
  \label{sec:background}
  
  \subsection{CHERI Protection}
  \label{sec:cheri}
  
  CHERI introduces capability-based pointer protection in
  HW~\CITE{2014-WoodruffWatson.etal}.
  Capabilities extend classical address
    pointers (i.e., plain integers) with access permissions (e.g., read,
  write and execute), address bounds and a \emph{valid}
  tag.
  Capabilities are typed to support restricting them to a special purpose, e.g., context switches. %
  Legal memory accesses require valid capabilities that hold corresponding
  permissions and whose bounds include the accessed address. %
  Instruction fetches are protected by the Program Counter Capability
  (PCC) which extends the normal PC with corresponding capability
  information. %
  
  Handling capabilities instead of plain integer pointers requires
  changes to the ISA as well as to the HW. %
  The ISA is extended by additional instructions that are needed for manipulating capabilities such as loading, storing or modifying them. %
  Legal capability modifications may only restrict but never expand the access permissions or address bounds. %
  Attempts to broaden the access rights result in
  clearing the \emph{valid} tag of the capability and therefore
  invalidating it for further use. %
  This important principle is called \emph{capability monotonicity}. %
  The HW implementation of the ISA changes includes adaptation of the register
  file to enable the storage of capabilities and augmentation of
  the memory locations with a tag bit
  indicating the integrity of the memory content. %
  Software (SW) running on a capability-enhanced processor starts with a set
  of initial capabilities provided by the architecture. %
  All other capabilities can be derived by SW according to the
  capability monotonicity property. %
  CHERI supports two modes: %
  \emph{pure capability} and \emph{hybrid}. %
  In hybrid mode memory accesses using plain ``legacy'' integer pointers are legal to support legacy code. %
  Pure capability mode only allows memory accesses using capabilities.
  
  \subsection{Related Work}
  \label{sec:related-work}
  
  Several works exist that verify security properties on
  the architecture definition of CHERI. %
  In~\CITE{2020-NienhuisJoannou.etal} the CHERI-MIPS ISA is formulated
  in the instruction set architecture description languages
  L3~\cite{2012-Fox} and SAIL~\cite{2019-ArmstrongBauereiss.etal}. %
  Security properties are then formally proven with theorem provers,
  e.g., Isabelle~\cite{2023-NipkowPaulson.etal}, on the ISA model. %
  Similarly,~\cite{2022-BauereissCampbell.etal}
  and~\cite{2023-GrisenthwaiteBarnes.etal} verify security properties on
  the capability-enhanced ARM Morello architecture specification. %
  The resulting ISA formalization helped to discover security issues and
  can be used to generate test cases for pre-silicon validation. %

  For both, CHERI-MIPS, as well as ARM
  Morello, ISA verification took a significant effort of three to four
  person-years. %
  The conducted proofs verify security
  aspects only for the architectural ISA model --- they do not verify security properties on an
  actual microarchitectural implementation of the ISA. %
  Proving a microarchitectural implementation secure requires additional
  verification steps and a sound translation of the ISA properties to
  microarchitectural properties of a specific HW implementation. %
  This may involve high manual effort and is prone to errors. %
  Furthermore, certain classes of vulnerabilities are not captured by
  the ISA model due to missing information, such as side channels based
  on timing. %
  Regarding transient execution side channel
  attacks,~\cite{2021-FuchsWoodruff.etal} shows that a CHERI-\RISCV{}
  HW implementation is vulnerable to several classes of Spectre
  attacks. %
  In addition, further
  analysis~\CITE{2021-Fuchs}
  has identified a vulnerability of CHERI systems to specific
  Meltdown-style attacks and transient privilege escalation. %
  Although this type of vulnerability can be mitigated by dedicated
  security mechanisms, other work has shown that even minor changes in
  implementation can (re-)open microarchitectural side
  channels~\CITE{2023-FadihehWezel.etal}.
  This drives the need for exhaustive security verification at the
  microarchitectural level. %

  Techniques for processor verification at the microarchitectural level
  have been proposed for various architectures.
  In \CITE{2016-ReidChen.etal} formal properties are machine-translated from the ARM ISA specification
  and checked on commercial ARM processors. %
  A verification framework is also proposed for x86,
  handling the complexity posed by a microcoded processors \CITE{2020-GoelSlobodova.etal}.
  The work of~\cite{2021-GaoMelham} verifies properties on CHERI
  Flute~\CITE{web-flute}, a HW 
  implementation of a capability-enhanced processor. %
  Here, the properties are translated manually from the SAIL
  specification of the CHERI-specific instructions in the CHERI-\RISCV{}
  ISA to SystemVerilog Assertions. %
  As discussed above, properties
  derived from the ISA specification may miss security vulnerabilities
  due to the semantic gap between the specification and the HW
  implementation. %
  In contrast, the methodology presented in this paper verifies global
  security properties based on a well-defined security objective
  (cf.~Sec.~\ref{sec:attacker-model}). %
  Our proofs cover all security aspects w.r.t.\ our security objective,
  including vulnerabilities that are not visible at the ISA level and
  therefore not captured by previous works. %

  As an alternative, complete formal
  verification~\cite{2014-UrdahlStoffel.etal} can be used to cover all
  possible operations within a HW implementation by
  operation-level properties. %
  For example, CS2QED~\cite{2020-DevarajegowdaFadiheh.etal} provides a highly
  automated and complete verification technique that can detect both
  single- and multiple-instruction bugs in in-order processor designs. %
  The drawback of this and related approaches with regards to security
  is that even complete functional verification cannot detect security
  vulnerabilities that are not caused by bugs in
  the implementation, such as side channels. %
  To cope with such threats, Unique Program Execution Checking
  (UPEC)~\cite{2023-FadihehWezel.etal} enables spec-free security
  verification based on formalized threat models. %
  The global security guarantee provided by UPEC depends on the
  underlying threat model. %
  So far, this verification methodology has not been extended to verify
  CHERI. %

  The work of~\CITE{2018-HuArdeshiricham.etal} introduces a HW verification approach based on information flow properties.
  Information flow tracking models are tailored to specific security properties, including confidentiality and integrity properties.
  However, application of this method requires a~priori knowledge of potential vulnerabilities.
  Therefore, the method is not exhaustive.

  \section{Formal Model}
  \label{sec:formal-model}
  
  \subsection{Security Objective}
  \label{sec:attacker-model}
  
  In this paper we assume a single-threaded capability-enhanced
  single-core processor with a memory for instructions and data. %
  We model the HW as a finite state machine (FSM),
  $(S, S_0, I, O, \lambda, \sigma)$, with a set of states~$S$, an
  initial state~$S_0$, a set of inputs~$I$ and outputs~$O$, a state
  transition function $\lambda : S \times I \to S$ and an output
  function $\sigma : S \times I \to O$. %
  We consider~$S$ to be valuations of the state
  variables and denote~$Z$ as the set of all state
  variables. %
  In the RTL model, $Z$
  corresponds to all state-holding elements, i.e., registers,
  microarchitectural buffers, caches, and memory locations. %
  We partition~$Z$ into two disjoint subsets: %
  
  \begin{itemize}
      \item~$P$: %
  containing all state variables in the processor;
      \item~$M$: %
  containing all memory locations. %
  \end{itemize}
  
  We further assume the processor to be executing (mutually distrusting) tasks and a trusted entity,
  such as an operating system, to securely manage context switches between tasks. %
  Context switches may be initiated by exceptions or interrupts or by dedicated control-flow transfers through specially typed capabilites.
  Each task has at its disposal a set of CHERI capabilities which
  grant access to a subset of the memory. %
  For each task running on the processor, we partition~$M$ into: %
  
  \begin{itemize}
      \item $M_\textit{pub}$: %
  denoting all public memory locations, i.e., all memory locations which are accessible with the attacker task's capabilities;
      \item $M_\textit{prot}$: %
  denoting all memory locations protected from the attacker task, i.e., all memory locations which are \emph{not} accessible with the attacker task's capabilities. %
  \end{itemize}
  
  Among the tasks, we assume an attacker with the objective to breach the memory isolation created by the CHERI capabilities. %
  In particular, the attacker tries to read protected data or execute protected instructions, i.e., violate \emph{confidentiality},
  or manipulate protected data, i.e., violate \emph{integrity}. %
  Confidentiality and integrity can be expressed as properties of the HW under SW constraints as given by capabilities. %
  We formulate these properties based on the strong and well-established notion of non-interference~\cite{1982-GoguenMeseguer}. %
  Non-interference defines \emph{high} and \emph{low} locations in the design and requires
  that the low locations are not affected by the high locations~\CITE{2010-ClarksonSchneider}. %
  We formulate a 2-safety non-interference computation tree logic (CTL) property for confidentiality (Eq.~\ref{eq:conf-non-interference})
  and integrity (Eq.~\ref{eq:int-non-interference}) of protected data. %
  We extend our notation of state variables by the~$\$$-operator to
  denote the valuation of a state variable set, i.e., the set of value assignments to the state variables. %
  
  For any task that can run on the system, i.e., for any choice of
  $M_\textit{prot}$ and $M_\textit{pub}$, it must hold: %
  \begin{equation}
  \label{eq:conf-non-interference}
  \begin{split}
      \textit{AG}\,(\,&\$M_\textit{pub} = \$M'_\textit{pub} \land \$P = \$P' \\
      &\rightarrow \textit{AG}\, (\,\$M_\textit{pub} = \$M'_\textit{pub} \land \$P = \$P'\,)\, )
  \end{split}
  \end{equation}
  
  \begin{equation}
  \label{eq:int-non-interference}
  \begin{split}
      \textit{AG}\,(\,&\$M_\textit{prot} = \$M'_\textit{prot} \rightarrow \textit{AG}\, (\,\$M_\textit{prot} = \$M'_\textit{prot}\,)\, )
  \end{split}
  \end{equation}
  
  In accordance to their 2-safety nature, both properties take two instances (primed$'$ and non-primed)
  of the HW model into consideration. %
  For the confidentiality property in Eq.~\ref{eq:conf-non-interference},
  the memory locations in $M_\textit{prot}$ are considered \emph{high} locations,
  while the locations accessible by the attacker $M_\textit{pub}$ and~$P$ are considered \emph{low} locations. %
  By assuming~$\$M_\textit{pub} = \$M'_\textit{pub}$ and~$\$P = \$P'$ we ensure that
  the corresponding state variables are equal between the two model instances. %
  The only difference between the two instances consists in state variables of $M_\textit{prot}$. %
  The property expresses that this difference never affects the \emph{low}
  locations, $M_\textit{pub}$ and~$P$, of the attacker task. %
  No protected information should ever become accessible to the attacker task. %
  The integrity property in Eq.~\ref{eq:int-non-interference} expresses interference in the opposite direction,
  by considering $M_\textit{prot}$ to be the locations that must not be influenced by
  the locations $M_\textit{pub}$ and~$P$. %
  In other words, the attacker task must never be able to modify protected information. %
  
  \subsection{Modeling Confidentiality and Integrity}
  \label{sec:model-conf-int}
  
  Proving the CTL properties of
  Eq.~\ref{eq:conf-non-interference} and
  Eq.~\ref{eq:int-non-interference} on realistic designs poses two major
  challenges. %
  The first challenge consists in modeling $M_\textit{pub}$ and
  $M_\textit{prot}$. %
  As defined in
    Sec.~\ref{sec:attacker-model}, the partitioning
  of all memory locations into $M_\textit{pub}$ and $M_\textit{prot}$ is
  determined by the capabilities available to the attacker task. %
  In the RTL model of the~HW, these capabilities are specified by
  configurations of address ranges, access permissions and additional
  information in capability registers and buffers. %
  This means, we need to express $M_\textit{pub}$ and $M_\textit{prot}$
  in terms of capabilities in the RTL. %
  
  The second challenge is posed by creating a proof that considers
  \emph{all} possible attacker tasks. %
  This requires considering all possible divisions of memory locations
  into $M_\textit{pub}$ and $M_\textit{prot}$, without rendering the
  proof unscalable. %
  
  Our key idea to overcome these challenges is the following: %
  Instead of modeling $M_\textit{pub}$ and $M_\textit{prot}$ explicitly by using state variables for memory locations,
  we model memory locations implicitly by their unique addresses. %
  This is possible because a violation of the two non-interference properties~Eq.~\ref{eq:conf-non-interference}~(confidentiality) and~Eq.~\ref{eq:int-non-interference}~(integrity)
  can only happen if the processor performs a memory access to $M_\textit{prot}$, assuming a correct memory implementation. %
  (Note that this is true even though $M_\textit{prot}$ does not appear in Eq.~\ref{eq:conf-non-interference}.) %
  As a result, we replace the consequent of the implication in the CTL properties from Eq.~\ref{eq:conf-non-interference} and
  Eq.~\ref{eq:int-non-interference} by checking if the processor
  performs a memory access to an address belonging to $M_\textit{prot}$. %
  We introduce the following macros, which are formulated using the processor's memory port signals: %
  
  \begin{itemize}
      \item \textit{read\_mem\_access}: %
  is true when the processor requests a read access to the memory. %
      \item \textit{write\_mem\_access}: %
  is true when the processor requests a write access to the memory. %
      \item \textit{mem\_addr}: %
  is the address for which the processor requests the memory access (read or write)
  \end{itemize}
  
  This leaves the challenge of covering all possible partitions of
  memory locations into $M_\textit{pub}$ and $M_\textit{prot}$ without
   enumerating all possible configurations explicitly. %
  We solve this problem by introducing a new \emph{symbolic address} for
  memory locations in~$M_\textit{prot}$, which is not connected to or
  constrained by the HW model and can thus freely be chosen by the
  solver of a property checking tool. %
  Furthermore, we introduce a new macro
  \textit{cheri\_protected}(\textit{symbolic\_addr}). %
  This macro constrains
  all registers and microarchitectural buffers holding capabilities to
  deny accesses to the symbolic address. %
  How \textit{cheri\_protected} is determined systematically is described
  in~Sec.~\ref{sec:method}. %
  The described modifications simplify the properties
  of~Eq.~\ref{eq:conf-non-interference}
  and~Eq.~\ref{eq:int-non-interference} to: %
  
  \begin{equation}
  \label{eq:conf-1-safety}
  \begin{split}
      \textit{AG}\,(\,&\textit{cheri\_protected} (\textit{symbolic\_addr}) \\
      & \rightarrow \,(\, \textit{read\_mem\_access}\, \rightarrow \,\textit{mem\_addr} \neq \textit{symbolic\_addr}\,)\, ) 
  \end{split}
  \end{equation}
  
  \begin{equation}
  \label{eq:int-1-safety}
  \begin{split}
      \textit{AG}\,(\,&\textit{cheri\_protected} (\textit{symbolic\_addr}) \\
      & \rightarrow \,(\, \textit{write\_mem\_access}\, \rightarrow \,\textit{mem\_addr} \neq \textit{symbolic\_addr}\,)\, ) 
  \end{split}
  \end{equation}
  
  Apart from the \emph{symbolic address}, the properties in
  Eq.~\ref{eq:conf-1-safety} and Eq.~\ref{eq:int-1-safety} have two
  notable differences compared to the non-interference properties
  of~Eq.~\ref{eq:conf-non-interference}
  and~Eq.~\ref{eq:int-non-interference}. %
  Firstly, by modeling memory locations as addresses, the properties no
  longer need to explicitly compare two design instances to detect a
  security violation. %
  This allows us to formulate the properties
  in~Eq.~\ref{eq:conf-1-safety} and~Eq.~\ref{eq:int-1-safety} as
  1-safety properties, proving the absence of any illegal memory
  access. %
  Secondly, the properties do not contain a universal quantifier, i.e.,
  $AG$, in the consequent of the implication. %
  Instead, the properties describe behavior without transitioning
  between states. %
  With regard to the RTL model, this means that the property formulates
  behavior in a single clock cycle. %
  Properties describing behavior spanning a finite number of clock
  cycles can be formulated as interval
  properties~\CITE{2014-UrdahlStoffel.etal}. %
  This includes the CTL properties in~Eq.~\ref{eq:conf-1-safety}
  and~Eq.~\ref{eq:int-1-safety}. %
  
  Interval properties can deliver an unbounded proof
  result due to their symbolic starting state. %
  In interval property checking (IPC), the property checker
  can fast-forward to any starting state
  violating the property without the need to unroll from reset. %
  The interval properties for our confidentiality and integrity security
  objectives are presented in Fig.~\ref{fig:conf-ipc} and
  Fig.~\ref{fig:int-ipc}. %

\begin{figure}[t]

    \begin{subfigure}{\linewidth}
        \begin{property}{}
1-safety-confidentiality: 
assume: 
  at t: cheri_protected(symbolic_addr);
prove: 
  at t: read_mem_access $\rightarrow$ mem_addr $\neq$ symbolic_addr;
        \end{property}
        \vspace{-4pt}
        \caption{Interval property for confidentiality} \label{fig:conf-ipc}
    \end{subfigure}
    \vspace{2pt}
    \begin{subfigure}{\linewidth}
        \begin{property}{}
1-safety-integrity: 
assume: 
  at t: cheri_protected(symbolic_addr);
prove: 
  at t: write_mem_access $\rightarrow$ mem_addr $\neq$ symbolic_addr;
        \end{property}
        \vspace{-4pt}
        \caption{Interval property for integrity} \label{fig:int-ipc}
    \end{subfigure}
    \vspace{2pt}

    \caption{1-safety non-interference interval properties} 
    \label{fig:ipc-prop}
\end{figure}

  \subsection{Proving Monotonicity}
  \label{sec:monotonicity}
  
  A successful proof of the interval properties
  in~Fig.~\ref{fig:conf-ipc} and~Fig.~\ref{fig:int-ipc} verifies that
  an attacker task cannot violate the confidentiality and integrity
  objectives of our threat model --- as long as the attacker task's
  capabilities comply with the \textit{cheri\_protected} macro. %
  Therefore, in order to achieve global proof validity, we need to prove
  that \textit{cheri\_protected} holds for the design during any
  attacker task execution. %
  
  We can easily prove this invariant by induction. %
  For proving the induction base,
  we have to show that the state space constrained by \textit{cheri\_protected} is reachable from reset. %
  (We omit the induction base property for reasons of space.) %
  The induction step is performed by the interval property depicted in Fig.~\ref{fig:mono-ipc}. %
  
\begin{figure}[t]
        \begin{property}{}
Monotonicity: 
assume: 
  at t:   cheri_protected(symbolic_addr);
prove: 
  at t+1: cheri_protected(symbolic_addr);
        \end{property}
    \caption{Interval property for monotonicity} 
    \vspace{-2mm}
    \label{fig:mono-ipc}
\end{figure}

  As introduced in Sec.~\ref{sec:model-conf-int}, the
  \textit{cheri\_protected}(\textit{symbolic\_addr}) macro consists of
  constraints for all registers or microarchitectural buffers that
  are accessible by the current task and that hold a capability. %
  Each constraint specifies that the respective capability does not
  grant access to the symbolic address, i.e., to any address
  pointing to a memory location in $M_\textit{prot}$. %
  The macro also includes constraints for capabilities loaded from and stored to memory by
  restricting the memory port signals in an analogous manner. %

  The property of Fig.~\ref{fig:mono-ipc} describes a
  monotonicity requirement similar to the one CHERI adheres to
  (cf.~Sec.~\ref{sec:background}). %
  This becomes apparent when expressing monotonicity with the
  notations used in this paper. %
  For the given disjoint sets of state variables, $P$,
  $M_\textit{pub}$, and $M_\textit{prot}$,
  it is required that no state transition can change capabilities
  to grant access to previously protected memory
  locations, i.e., moving them from $M_\textit{prot}$ to~$M_\textit{pub}$. %
  We call this notion \emph{weak capability monotonicity}. %
  
  We denote:
  \begin{itemize}
      \item $M^t_\textit{prot}$ as the set of protected memory locations \emph{before} a state transition and
      \item $M^{t+1}_\textit{prot}$ as the set of protected memory locations \emph{after} a state transition.
  \end{itemize}
  Weak capability monotonicity requires that $M^{t+1}_\textit{prot} \subseteq M^t_\textit{prot}$, i.e., $M^{t+1}_\textit{prot} \setminus M^t_\textit{prot} = \emptyset$. %
  In other words, state transitions never add to the set $M_{prot}$. %
  The connection to the interval property in~Fig.~\ref{fig:mono-ipc} is apparent when considering a case where monotonicity
  is violated, i.e., $M^{t+1}_{prot} \setminus M^t_{prot} = M^{t+1}_\textit{vio} \neq \emptyset$. %
  In such a case, the property checker would create a counterexample and
  choose a symbolic address pointing to a memory location in $M^{t+1}_\textit{vio}$. %
  
  There is a special case which we have omitted in~\reffig{mono-ipc}
  for clarity of presentation, but which nevertheless must be treated
  in the practical implementation of the property. %
  The case occurs when there is a context switch to another task and
  monotonicity ``ends'' for the considered current task. %
  In CHERI, the task uses a specially typed capability to perform a
  jump to an address in $M_\mathit{prot}$, handing over control of the
  processor to the trusted operating system (OS), as specified in our security objective
  (cf.~Sec.~\ref{sec:attacker-model}). %

  Weak capability monotonicity is a weaker version of CHERI's capability monotonicity. %
  While capability monotonicity forbids individual capabilities to grow to more access
  rights, e.g., larger bounds, weak capability monotonicity only asks whether
  the set of capabilities controlled by a task can obtain more access rights overall. %
  For our formulation of properties, weak capability monotonicity is a sufficiently strong invariant. %
  It is easier to formulate than capability monotonicity and it synergizes
  well with the concept of the symbolic address. %
  A notion similar to weak capability monotonicity called \emph{reachable capability monotonicity} is discussed
  in~\CITE{2020-NienhuisJoannou.etal} for the ISA model. %

  \subsection{Reformulating UPEC for CHERI}
  \label{model-upec}
  
  The three properties in Fig.~\ref{fig:conf-ipc}, \ref{fig:int-ipc} and \ref{fig:mono-ipc} are sufficient to prove non-interference and consider every access to protected data a security violation. %
  This is reasonable for integrity, i.e., every write access to protected data can be considered a security breach. %
  However, for confidentiality, this can be too conservative. %
  Preventing every read access to protected memory is a sufficient condition for security while preventing the actual propagation to an attacker-visible state is both, a necessary and sufficient condition. %
  
  It may happen that accesses to protected data are executed in parallel
  with an access control check evaluating legality of the access. %
  In case it is not legal, an exception is raised and the loaded data is
  discarded. %
  However, if the discarded data leaves a footprint in the
  microarchitecture, e.g., in a data cache, and the timing of any
  subsequent operations in the processor depends on this footprint, then
  a timing side channel violating confidentiality exists. %
  In accordance with~\cite{2023-FadihehWezel.etal}, we call the
  corresponding attack Meltdown-style attack. %
  Such vulnerabilities can be detected using the Unique Program
  Execution Checking (UPEC)~\cite{2023-FadihehWezel.etal} verification
  method. %
  
  UPEC was introduced to detect vulnerabilities to Transient
  Execution Attacks in processors. %
  UPEC verifies security using a 2-safety property running on a miter
  model. %
  This miter instantiates two instances of the design under verification
  and constrains the public inputs as well as the symbolic initial state
  to be equal. %
  Hence, any difference between the two instances must originate in the
  difference caused by the protected data. %
  UPEC distinguishes between propagation of protected data to
  microarchitectural buffers (\emph{\PALERT{}}) and actual leakage of
  the protected data to the architectural state (\emph{\LALERT{}}). %
  
  In the following, we adapt UPEC to designs featuring CHERI protection and use it in addition to proving the properties of the previous sections. %
  This serves two purposes. %
  Firstly, we remove the above-mentioned conservatism of Eq.~\ref{eq:conf-non-interference}
  by taking into account whether or not protected data can propagate to an attacker-visible state. %
  Secondly, we detect also Meltdown-style side channels. %
  
  As a first step, we further divide the set of \emph{microarchitectural state variables}~$P$ into the two disjoint subsets $P_\textit{arch}$ and $P_\textit{\mbox{$\mu$-arch}}$: %
  
  \begin{itemize}
  \item $P_\textit{arch}$: %
    containing all \emph{architectural} state
    variables in the processor, i.e., all registers visible to the
    attacker task; %
  \item $P_\textit{\mbox{$\mu$-arch}}$: %
    containing all but the architectural state
    variables in the processor, i.e., all microarchitectural
    registers and buffers \emph{not} visible to the attacker task. %
  \end{itemize}
  
  With this addition to our model, we formulate UPEC as a 2-safety CTL property. %
  
  For any task that can run on the system, i.e., for any choice of
      $M_\textit{prot}$ and $M_\textit{pub}$, it must hold: %
      \begin{equation}
        \label{eq:conf-upec}
  \begin{split}
    \textit{AG}\,(\,&\$P_\textit{arch} = \$P'_\textit{arch}
                      \land \$P_\textit{\mbox{$\mu$-arch}} = \$P'_\textit{\mbox{$\mu$-arch}}
                      \land \$M_\textit{pub} = \$M'_\textit{pub} \\
                    &\rightarrow \textit{AG}\, (\,\$P_\textit{arch} = \$P'_\textit{arch} \,)\, )
  \end{split}
  \end{equation}
  
  The property only allows a difference between the two model instances in the protected memory locations $M_\textit{prot}$;
  all other state variables are equal between the two instances. %
  The property then verifies that the architectural state variables $P_\textit{arch}$, i.e., the attacker-visible part of the processor, are never influenced by contents of $M_\textit{prot}$. %
  In other words, the protected information is allowed to enter the processor core,
  but must not influence the architectural state variables. %
  
  The property in Eq.~\ref{eq:conf-upec} takes the same perspective as
  the 1-safety property in Eq.~\ref{eq:conf-1-safety}. %
  The
  parititioning of memory locations into
  $M_\textit{pub}$ and $M_\textit{prot}$ is specified for a
  generic attacker task running on the processor. %
  This is an important
  difference compared to the UPEC property presented
  in~\cite{2023-FadihehWezel.etal}. %
  In~\cite{2023-FadihehWezel.etal},
  UPEC is formulated globally for a design and independently of the
  running task. %
  As a result, our UPEC formulation does not require any sound
  approximation excluding the case that protected data is leaked
  intentionally (cf.~equivalence of \emph{architectural observations} in~\cite{2023-FadihehWezel.etal}). %
  
  Analogously to~\cite{2023-FadihehWezel.etal}, we prove
  Eq.~\ref{eq:conf-upec} by an inductive proof procedure based on the
  interval property of Fig.~\ref{fig:upec-ipc} formulated for a time window of $k$ clock cycles. %
  To this end, we model~$P$ and~$P'$ as two instances of the processor, but model
  $M_\textit{pub}$ and $M'_\textit{pub}$ by their addresses, similar as
  for the 1-safety properties in Eq.~\ref{eq:conf-1-safety}. %
  This allows us to use the same macro
  \textit{cheri\_protected} and the symbolic address to model
  protection by CHERI capabilities. %

\begin{figure}[t]
\begin{property}{}
UPEC-CHERI: 
assume: 
  at t:   cheri_protected(symbolic_addr);
  at t:   $\$M_\textit{pub} = \$M'_\textit{pub} \land \$P = \$P'$;
prove: 
  at t+k: $\$P_\textit{arch} = \$P'_\textit{arch}$;
\end{property}
    \caption{UPEC property for CHERI} 
    \vspace{-2mm}
    \label{fig:upec-ipc}
\end{figure}

  \section{Formal Verification Flow}
  \label{sec:method}
  
\begin{figure}
    \centering
    \includegraphics{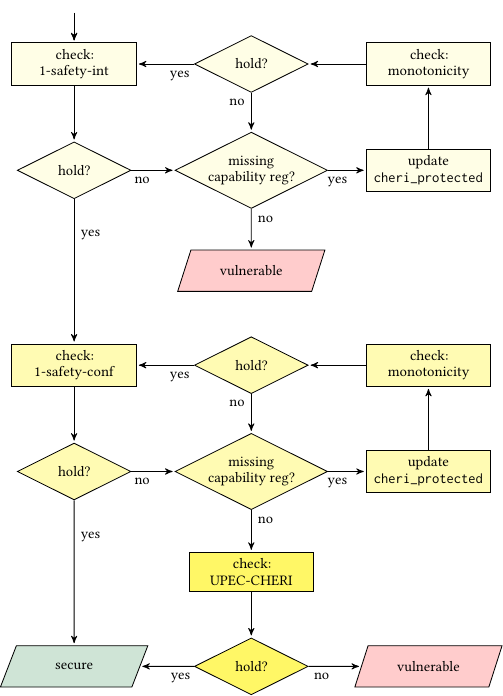} 
    \caption{VeriCHERI Verification Flow}
    \label{fig:formal_flow}
  \end{figure}

  We propose VeriCHERI, a novel formal verification flow based on the
  interval properties introduced in Sec.~\ref{sec:formal-model}. %
  The verification flow exhaustively detects all violations to the
  security objectives formulated in~Sec.~\ref{sec:attacker-model} and
  provides formal security guarantees after removing all detected
  vulnerabilities in the design. %
  As a part of the flow, the macro
  \textit{cheri\_protected} is determined
  iteratively. %
  The flow is illustrated in Fig.~\ref{fig:formal_flow}. %
  
  VeriCHERI starts by proving the 1-safety integrity property
  (cf.~Fig.~\ref{fig:int-ipc}). %
  If the property fails, a counterexample is returned and the
  verification engineer must decide whether it is a
  false or true one. %
  A false counterexample can occur if the macro
  \textit{cheri\_protected} misses to constrain a register or buffer
  holding a capability such that, spuriously, access to protected data
  is granted. %
  This reflects behavior that is unreachable for the current settings of
  $M_\textit{prot}$. %
  We remove this false counterexample by refining
  \textit{cheri\_protected} with a constraint on the concerned register or
  buffer, as described in \refsec{monotonicity}. %
  
  In the next step, by proving the property in
  Fig.~\ref{fig:mono-ipc}, we verify that the updated macro
  \textit{cheri\_protected} remains an invariant, i.e., weak capability
  monotonicity is preserved. %
  If monotonicity fails, the verification engineer must inspect the new
  counterexample to modify or further refine
  \textit{cheri\_protected}. %
  
  The verification engineer detects a true security vulnerability if the
  counterexample shows the execution of a task that creates a capability
  granting access to protected data or that accesses protected data
  without a legal capability. %
  This reflects behavior that is reachable for the current
  settings of $M_\textit{prot}$. %
  
  In case the monotonicity property holds, the integrity property can be
  run again with the updated \textit{cheri\_protected} macro. %
  This step is repeated until a vulnerability is detected or the
  1-safety integrity property holds successfully. %
  
  Once the 1-safety integrity property holds, the confidentiality
  property (cf.~Fig.~\ref{fig:conf-ipc}) is checked. %
  In principle, the same iterative procedure as discussed above must be
  followed. %
  However, in practice, (unless there is a capability used only in read
  accesses that is otherwise not accessible by the task), no iteration
  to update \textit{cheri\_protected} is needed. %
  
  If both 1-safety properties hold, the design is guaranteed to be free
  of vulnerabilities violating our security objective. %
  However, if the 1-safety-confidentiality property detects
  a vulnerability, we know that protected data is accessed. %
  In this case UPEC-CHERI~(cf.~Fig.~\ref{fig:upec-ipc}) is used to further investigate the counterexample and check
  whether the protected data indeed propagates to some attacker-visible
  state. %

  \section{Case study}
  \label{sec:case-study}
  
  We conducted a case study on the CHERIoT Ibex processor implementing the CHERIoT architecture~\CITE{2023-AmarChisnall.etal}. %
  CHERIoT is a variant of CHERI that aims to provide memory safety for embedded Internet of Things (IoT) real-time devices based on a compartmentalization model. %
  Compartments are mutually distrusting address spaces containing code and data, isolated from each other by capabilities. %
  All code is run in pure capability mode (cf.~\refsec{cheri}). %
  In CHERIoT, threads are individual instances of execution, owning a stack pointer and register file contents, and
  are scheduled to run in compartments. %
  Compartment and context switches are performed by designated procedures administered by a trusted OS. %
  Capabilities stored in registers and memory are compressed
  to address the resource limitations for embedded IoT devices. %
  
  All experiments were conducted on a workstation PC featuring an Intel
  i9-13900k with 128 GB of RAM running Linux and using the commercial property
  checker OneSpin 360 DV by Siemens EDA. %
  
  \subsection{Applying VeriCHERI}
  
  \begin{table*}
  \centering
  \begin{tabular}{lccrrl}
  	
    \hline &&&&& \\ [-2.1ex]
    \hline &&&&& \\ [-1.5ex]
    \textbf{Property} & \textbf{Iteration}  & \textbf{Result} & \textbf{Runtime} & \textbf{Memory} & \textbf{Description} \\ [0.5ex]
    \hline &&&&& \\ [-1.5ex]
      
    1-safety-integrity & 1 & \textcolor{red}{fail}   & < 1 min & 4.3 GB & \textit{Bug}: setup guide specification of protection enable pin \\
              & 2 & \textcolor{red}{fail}            & < 1 min & 4.7 GB & \textit{Bug}: capability stores across capability bounds  \\
              & 3 & \textcolor{green}{hold}          & 7 min   & 4.8 GB & -  \\ [0.5ex]
                
    \hline &&&&& \\ [-1.5ex]
      
    Monotonicity & 1-9  & \textcolor{red}{fail}   & $\leq$ 1 min & 4-5 GB & Missing capability register or pipeline buffer  \\
                 & 10    & \textcolor{green}{hold} & 15 min & 6.2 GB & -  \\ [0.5ex]
                   
    \hline &&&&& \\ [-1.5ex]
      
    1-safety-confidentiality               &&&&& \\
    $\longrightarrow$ data         & 1 & \textcolor{green}{hold} & 7 min   & 7.3 GB & -  \\
    $\longrightarrow$ instructions & 1 & \textcolor{red}{fail}   & < 1 min & 4.8 GB &  Instruction fetched from outside PCC bounds \\ [0.5ex]
      
    \hline &&&&& \\ [-1.5ex]
      
    UPEC-CHERI & 1 & \textcolor{red}{fail}   & 31 min & 3.7 GB &  \textit{Side channel}: exception timing depends on fetched data \\
               & 2 & \textcolor{green}{hold} & 18 min & 6.3 GB &  - \\ [0.5ex]
                   
    \hline &&&&& \\ [-2.1ex]
    \hline &&&&& \\ [-1ex]
    
  \end{tabular}
  \caption{CHERIoT Properties}
  \label{tab:properties}
\end{table*}
  We applied the formal verification flow presented
  in~Sec.~\ref{sec:method} to the CHERIoT Ibex RTL design~\CITE{web-cheriot}. %
  The results are summarized in Tab.~\ref{tab:properties}.
  In accordance with the flow, we started with the setup and proof of
   the 1-safety integrity property (cf.~Fig.~\ref{fig:int-ipc}). %
  In the first iteration, the property failed due to the initially incomplete specification
    of the \textit{cheri\_protected} macro. %
   We updated the \textit{cheri\_protected} macro in several iterations
  until we successfully verified the monotonicity property
  (cf.~Fig.~\ref{fig:mono-ipc}). %
  Overall, the iterations detected 40 locations in the core which hold
  capabilities, including the 31~capabilities in the general purpose
  register file, seven special control registers, the PCC 
  and one pipeline
  buffer in the writeback stage.
  
  During this process, we detected two security bugs. %
  The first bug was related to a primary input pin responsible for
  globally enabling CHERI protection. %
  In the setup guide, the pin was specified as active low, but
  implemented as active high in the RTL. %
  We reported the bug to the CHERIoT developer team who updated the setup guide
  accordingly. %
  
  For the second bug, the property checker produced a counterexample
  where the first memory access of a capability store lies within the
  allowed address bounds, but the second memory access goes
  outside the bounds. %
  On capability loads and stores, the bounds
  were only compared to the address of the first memory access and
  failed to prohibit the second address. %
  During our experiments, this bug was discovered independently also by the
  development team and fixed by checking whether the address of the second
  memory access, i.e., the address of the first access incremented by~4, 
  also lies within bounds. %
  
  After successfully proving the
  1-safety-integrity property, we proceeded to checking the
  1-safety-confidentiality property (cf. Fig. \ref{fig:conf-ipc}). 
  For convenience, we split the property into two sub-properties, one
  property considering the data memory interface and one property for
  the instruction memory interface. %
  While the data memory property held, the instruction memory property
  failed and produced a counterexample. %
  The counterexample describes a scenario where instructions
  are fetched outside of legal capability bounds. %
  We ran the UPEC-CHERI property (cf.~Fig.~\ref{fig:upec-ipc}) to
  further investigate this problem. %
  
  UPEC detected a vulnerability to a potential Meltdown-style attack. %
  In this particular case, an attacker tries to jump to an address
  pointing to a memory location in $M_\textit{prot}$. %
  This starts a race with an exception which ultimately flushes the
  pipeline before the instruction is executed. %
  However, depending on the content of two specific bits of the fetched
  data, the pipeline flush is delayed. %
  By measuring the (overall) execution time, an attacker can probe the
  two bits for an arbitrary protected address. %
  In addition, the content of the two bits is reflected in an offset of a performance counter.
  While, in this particular case, the detected timing side channel can only leak a small amount of information per probed memory word, the example demonstrates that such timing side channels may well
  be present even in small processors such as CHERIoT Ibex, which can be a significant risk in general. Exhaustive    
  formal verification targeting such
  vulnerabilities is required for creating
  trust. %
  We reported this vulnerability and received  confirmation from the
  development team. %
  
  We applied a conservative fix that prevents all illegally fetched
  instructions to enter the core. %
  With this fix, the UPEC property holds. %
  
  The runtimes of the individual properties range from a few seconds to 31 minutes.

  \subsection{Invariants as Reusable Verification IPs}
  
  Verifying unbounded interval properties may produce false counterexamples.
  This happens when the property checker assumes unreachable states as symbolic starting states.
  If a false counterexample is identified,
  invariants can be formulated to exclude such unreachable states from consideration.
  The invariants are then proven by induction.
  This is standard practice in IPC, as supported by commercial property checking tools.
  
  While running the formal verification flow we formulated several invariants.
  Most of the formulated invariants are not specific to CHERI and exclude unreachable states
  in microarchitectural buffers such as control FSM registers.
  The formulated invariants were easy to prove and required only little manual effort.
  These invariants are implementation-specific and are, most likely, of limited use for another CHERIoT (or CHERI) processor.
  However, our flow of Fig.~\ref{fig:formal_flow} produces also two invariants that bear significance beyond this case study.
  
  The first one is the weak capability monotonicity invariant introduced in \refsec{monotonicity}.
  The invariant constrains a set of capability registers and buffers holding capabilities to
  deny accesses to protected data. 
  While the locations of these registers and buffers have to be re-determined for another processor, the constraints themselves can be reused.
  Depending on the similarity in register layout, the constraints require minimal to no modifications. 
  
  The second invariant worth reporting is necessary to accommodate for
  CHERIoT's compression scheme used to store capabilities in registers and memory.
  It consists of multiple conditions describing legal configurations
  of the capabilities in the system.
  The CHERIoT architecture requires the processor HW to enforce a similar set of conditions for 
  compressed capabilities. This set is called \emph{representability}~\cite{2023-AmarChen.etal}.
  Most of the complexity in formulating the invariant comes from the interdependence
  between the compressed capability bounds and the corresponding addresses.
  However, the formulation of the invariant benefits from a synergy with weak capability monotonicity.
  We specify an additional constraint for each register or buffer holding a compressed capability.
  In essence, it can be formulated as a function for any such register or buffer identified during the monotonicity proof.
  
  Both invariants are formulated in SVA and are published open-source~\cite{web-vericheri}.
  
  \subsection{Discussion}
  
  Although VeriCHERI  uses an abstract security objective as starting point of
  our verification procedures, the resulting guarantees can be related
  directly to the CHERI security requirements. %
  For the CHERIoT architecture used in our case study, mutually
  distrusting compartments can be modeled by assuming
  all memory locations within a compartment as $M_\textit{pub}$ and all memory locations
  outside the compartment as $M_\textit{prot}$ (cf.~Sec.~\ref{sec:attacker-model}).
  A task in our abstract threat model corresponds to a thread executing code in a compartment.
  The task's execution ends upon a context or compartment switch.
  
  In industrial practice, we intend VeriCHERI to be used in concert with established verification flows for functional correctness based on simulation or
  (non-complete) formal verification.
  We think that augmenting such
  existing flows with formal security guarantees is a ``sweet spot'' for VeriCHERI. 

  The manual effort involved in this case study approximates to two
  person months.  This number includes the development and refinement of
  the monotonicity and representability invariants.  Since they are
  reusable, this will significantly reduce effort in future applications
  of VeriCHERI.

\section{Conclusion}
\label{sec:conclusion}

We present VeriCHERI, a formal security verification method tailored
to CHERI processors.
It exhaustively detects confidentiality and integrity vulnerabilities
w.r.t. a global security objective based on non-interference.
In a case study on the CHERIoT Ibex core, VeriCHERI detected three security bugs,
including a vulnerability to a potential Meltdown-style timing side channel attack.
Such a vulnerability does not violate functional correctness and is thus not detected by previous verification methods based on verifying an ISA specification.
We also demonstrated that VeriCHERI is scalable and feasible in terms of manual effort.
Future work will apply VeriCHERI to more CHERI designs, including out-of-order processors.

\begin{acks}
    This work was supported partly by Bundesministerium f\"ur Bildung und Forschung Scale4Edge, grant no. 16ME0122K-16ME0140+ 16ME0465, by Intel Corp., Scalable Assurance Program, by Siemens EDA and by Deutsche Forschungsgemeinschaft SPP NanoSecurity, reference number KU 1051/11-2.
    We thank the CHERIoT Ibex development team for their valuable feedback. 
\end{acks}

\bibliographystyle{ACM-Reference-Format}
\bibliography{refs3}

\end{document}